Radiation Rates for Low Z Impurities in Edge Plasmas

R. Clark[1], J. Abdallah[1], and D. Post[2]

[1]Los Alamos National Laboratory, Los Alamos, NM

[2]ITER Joint Central Team, San Diego, CA

Abstract

The role of impurity radiation in the reduction of heat loads on divertor plates in present experiments such as DIII-D, JET, JT-60, ASDEX, and Alcator C-Mod, and in planned experiments such as ITER and TPX places a new degree of importance on the accuracy of impurity radiation emission rates for electron temperatures below 250 eV for ITER and below 150 eV for present experiments. We have calculated the radiated power loss using a collisional-radiative model for Be, B, C, Ne and Ar using a multiple configuration interaction model which includes density dependent effects, as well as a very detailed treatment of the energy levels and meta-stable levels. The "collisional radiative" effects are very important for Be at temperatures below 10 eV. The same effects are present for higher Z impurities, but not as strongly. For some of the lower Z elements, the new rates are about a factor of two lower than those from a widely used, simpler average-ion package (ADPAK)[1, 2] developed for high Z ions and for higher temperatures. Following the approach of Lengyel [3, 4] for the case where electron heat conduction is the dominant mechanism for heat transport along field lines, our analysis indicates that significant enhancements of the radiation losses above collisional-radiative model rates due to such effects as rapid recycling and charge exchange recombination will be necessary for impurity radiation to reduce the peak heat loads on divertor plates for high heat flux experiments such as ITER.

I. Introduction

Impurity radiation from low Z intrinsic impurities such as Be, B and C, and from other impurities such as Ne and Ar specifically introduced to enhance impurity radiation losses, can be an important factor in reducing the heat loads on the divertor plates in many present experiments and most planned experiments. Accurate knowledge of the radiation rates of these impurities is therefore essential both for an accurate assessment of the feasibility of this approach and for the analysis of present experiments in which impurity radiation in the plasma edge and divertor plasma plays an important role. The focus of most previous calculations of impurity radiation rates has been for the plasma core with high temperatures. Although there have been some detailed calculations for C and O, detailed calculations are needed for other elements including Be, B, Ne and Ar which are important for the plasma edge.

To estimate the appropriate temperature range, the maximum temperature near the separatrix near the plasma mid-plane can be expressed as a function of the parallel heat flux, $Q_\parallel$, (Equation 1), the safety factor, $q_{\psi 95\%}$, and the major radius, R.

$$Q_\parallel \approx \frac{P_{heat}}{A_\perp}, \quad where \quad A_\perp = \frac{2 \times 2\pi a \sqrt{\frac{1+\kappa^2}{2}} \, \delta}{q_{\psi 95\%}} \tag{1}$$

where $\delta$ is the radial decay length of the power in the plasma edge beyond the separatrix, a is the plasma minor radius, $P_{heat}$ is the heating power, $q_{\psi 95\%}$ is the MHD safety factor and $\kappa$ is the plasma elongation. Assuming that $T_{e-div} \ll T_{e-separatrix}$, so that $T_{e-div}$ can be neglected, we can estimate the temperature of the plasma edge by integrating the heat flux transported by electron conduction along the field lines in the scrape-off layer from the main plasma edge at the separatrix to the divertor plate (Eq. 2) (using $\ln \Lambda \approx 10$).

$$Q_\parallel = -\kappa_o T_e^{2.5} \frac{\partial T_e}{\partial x} = -\frac{2}{7} \kappa_o \frac{\partial T_e^{3.5}}{\partial x} \quad \therefore \quad T_{es}^{3.5} - T_{ediv}^{3.5} = \frac{1}{\kappa_o} \int_s^{div} Q_\parallel dx \approx \frac{7}{2} 1.4 \pi q R \frac{Q_\parallel}{\kappa_o}$$

$$T_{es} \approx 82.2 \, eV \left( Q_\parallel (GW/m^2) \, R(m) \, q \, Z_{eff} \right)^{2/7} \therefore where \, \kappa_o \approx \frac{3.1 \times 10^9}{Z_{eff} \ln \Lambda} \left( \frac{erg}{cm \, s \, eV^{3.5}} \right) \tag{2}$$

For present experiments such as DIII-D, the predicted maximum electron temperatures (for $\delta \sim 1$ cm) range from 80 to 120 eV, and for ITER they may be as large as 315 eV (Table 1).

Table 1 Typical edge temperatures, connection lengths, and parallel heat fluxes

|  | $P_\alpha$(MW) | $Q_\parallel$(GW/m$^2$) | $A_\perp$(m$^2$) | L (m) | R (m) | $T_s$(eV) |
|---|---|---|---|---|---|---|
| DIII-D 5 MW | 5 | 0.12 | 0.043 | 22 | 1.67 | 80 |
| DIII-D 20 MW | 20 | 0.47 | 0.043 | 22 | 1.67 | 120 |
| ITER 1.5 GW | 240 | 1.5 | 0.16 | 100 | 8.00 | 260 |
| ITER 3 GW | 480 | 3 | 0.16 | 100 | 8.00 | 315 |

These temperatures are low compared to those in the plasma core (~5-15 keV) and in the plasma edge several centimeters inside of the separatrix (~500 eV to 2 keV). Density dependent effects are potentially important because the plasma density in the divertor can be larger than the plasma core. Pressure balance along the field lines implies that as the plasma temperature decreases due to impurity radiation losses, the density rises. A temperature decrease from 100 eV to ~ 5 eV would increase the density of the radiating region from ~$5 \times 10^{19}$ m$^{-3}$ to $10^{21}$ m$^{-3}$. The temperatures for DIII-D are somewhat lower than measured on the experiment, probably due to a somewhat smaller $\delta$ resulting in a higher $Q_\parallel$. Ln $\Lambda$ varies from 8 to 13 as well in the range of interest. Nonetheless, the temperatures below can be used to assess the general range of conditions likely to be important.

Motivated by these considerations, we have performed detailed calculations of the radiation rates for Be, C, Ne and Ar using a multiple configuration interaction model which includes density dependent effects, as well as a very detailed treatment of the energy levels and meta-stable levels. The density is potentially important because high densities can cause collisional interruption of radiative decay process and suppression of di-electronic recombination and reduce the radiation emission.

The calculations of radiated power loss are based on the system of computer codes developed at Los Alamos [5, 6, 7]. For the light elements, Be, B, and C, calculations have been carried out in great detail using distorted wave theory for many of the collisional excitation cross sections. Due to the complexity of the systems, the Ne and Ar calculations were not done in as much detail. The radiated power loss calculations for Be, B and C are probably the most comprehensive calculations performed to date.

The effect of the changes in the radiation rates on the efficiency of cooling the divertor plasma can be estimated by using a model due to Lengyel[3] which balances electron heat conduction along the field lines with radiation losses to calculate the parallel heat flux for a given upstream density and overall impurity fraction that can be radiated from the flux tube.

Details of the models and assumptions used are given in Section II. In Section III results are presented for radiated power loss and radiation effectiveness. Conclusions are given in Section IV.

II. Model Description and Assumptions

The calculations of radiated power loss presented here were performed using the atomic physics codes developed at Los Alamos. The atomic structure calculations use the Hartree-Fock method developed by Cowan [8]and expanded upon by Clark and Abdallah[9] in the CATS computer code. The CATS code calculates energy levels, oscillator strengths, and plane wave Born (PWB) electron impact excitation cross sections. All of the atomic structure information is stored on a file for use by other codes. The ACE code[7] uses the structure information to calculate distorted wave (DW) cross sections for electron impact. Since the DW calculations are much more time consuming, the PWB cross sections are used for most transitions. DW calculations are performed on transitions from the ground state to all higher states. All ionization processes are calculated by the GIPPER code. The processes included are electron impact ionization, photoionization, and autoionization. The electron impact ionization makes use of the scaled hydrogenic cross sections which has compared well with distorted wave calculations [9] . The photoionization cross sections are necessary to obtain the radiative recombination cross sections by the principle of detailed balance. A large number of autoionizing states are included in the calculations. Dielectronic recombination is obtained by detailed balance. The atomic physics codes can be run in either a configuration average mode or in the fine structure levels mode. The configuration average mode calculates quantities for pure configurations of the electrons i.e.. from the set of nl and occupancies, without any coupling of angular momentum.  In the fine structure mode, the spin and orbital angular momenta of all the electrons are coupled vectorially to give rise to total spin, total orbital, and total angular momentum quantum numbers. Mixing of states with differing configurations, spin, and orbital quantum numbers is allowed in the fine structure mode. In the present calculations, the fine structure mode was used for Be, B, and C. Ne and Ar calculations were done in the configuration average mode because of the complexity of those systems. In the case of C, configuration average calculations were also performed for comparison purposes. In general, the calculated radiated power loss is less when done in the fine structure mode. This is due to the separation of spin states. The metastable states can be collisionally deexcited which means that less energy is available to radiation.

In all of the present calculations configurations through n=5 were included. For the light elements DW electron impact cross sections were calculated for all transitions from the lowest configuration. PWB cross sections were used for all other transitions. This includes transitions between all excited states. For Ne and Ar only PWB cross sections were used. For C, comparisons were made of calculations with PWB cross sections versus calculations with the ground state excitation cross sections replaced with DW calculations.

Because "closed shell" configurations such as fully stripped, He-like, Ne-like, etc. ions do not radiate strongly, the temperature range over which the impurities radiate can be very limited, especially for low Z impurities. Since the power must be transported to the region of the plasma with the temperature at which the impurities radiate by heat conduction which requires a temperature gradient, the range in $T_e$ for a given impurity may not be large enough to obtain adequately large radiation losses. Following Lengyel [3], the maximum power that can be radiated can be calculated assuming pressure balance along the field lines (equation 3).

$$\frac{\partial Q_{\parallel}}{\partial x} = -n_e n_z L_Z(T_e) \quad \vdots \quad Q_{\parallel} = -\kappa_o T_e^{2.5} \frac{\partial T_e}{\partial x} \quad \vdots \quad p_e = n_e T_e \quad \Rightarrow$$

$$Q_{\parallel} \frac{\partial Q_{\parallel}}{\partial x} = -n_e n_z L_Z Q_{\parallel} \approx \kappa_o n_e n_z L_Z T_e^{2.5} \frac{\partial T_e}{\partial x} \quad \Rightarrow$$

$$\frac{1}{2} \frac{\partial Q_{\parallel}^2}{\partial x} \approx \frac{p_e^2}{T_e^2} \kappa_o f_Z L_Z T_e^{2.5} \frac{\partial T_e}{\partial x} \approx p_e^2 \kappa_o f_Z L_Z T_e^{0.5} \frac{\partial T_e}{\partial x} \quad \Rightarrow$$

$$\frac{1}{2} d Q_{\parallel}^2 \approx p_{es}^2 \kappa_o f_z L_Z T_e^{0.5} dT_e \quad \Rightarrow \frac{\Delta Q_{\parallel}}{n_{es}\sqrt{F_z}} \approx \sqrt{2\bar{\kappa}_o T_{es}^2 \int_0^{T_{es}} L_Z(T_e) T_e^{0.5} dT_e}$$

$$where \; \kappa_o \approx \frac{3.1 \times 10^9}{Z_{eff} \ln \Lambda} \left( \frac{erg}{cm \; s \; eV^{3.5}} \right) and \; F_z(f_Z) = \frac{f_Z(\%)}{Z_{eff}} = \frac{f_Z(\%)}{1 + .01 f_Z(\%) Z(Z-1)}$$

$$with \; \bar{\kappa}_o \equiv \kappa_o Z_{eff}, \; n_e(cm^{-3}), \; T_e \; (eV), \; L_Z(\,ergs \; cm^3 \, s^{-1}) \; and \; \Delta Q_{\parallel}\left(\frac{ergs}{s \, cm^{-2}}\right) \tag{3}$$

$$In \; practical \; units: \frac{\Delta Q_{\parallel}\left(\frac{GWatts}{m^2}\right)}{n_{es}\left(10^{20} m^{-3}\right)\sqrt{F_z}} \approx 2.5 \times 10^5 \sqrt{T_{es}^2 \int_0^{T_{es}} L_Z(T_e) T_e^{0.5} dT_e}$$

where $p_s$, $n_s$, and $T_s$ are the pressure, density and temperature at the mid-plane, $f_Z$ is the impurity fraction in percent, $Q_{\parallel}$ is parallel heat flux, $L_Z$ is the radiation emissivity, $n_e$ is the electron density, and $n_Z$ is the impurity density. The dependence on field line length is implicit in the present treatment through $T_s$ which depends on the field line length. This result depends on electron heat conduction being the dominant energy transport mechanism along the field lines. "Flux-limited" heat transport retains an explicit dependence on the field line length.

## III. Results

Several different approximations available for calculation of radiated power loss for an electron density of $10^{12}$ cm$^{-3}$ are compared in Figure 1. In the figure the solid curve with the triangles was calculated using the configuration average mode and PWB cross sections for all transitions. The solid curve with squares was also done in the configuration average mode with PWB cross sections for all the excited to excited transitions, but using DW cross sections for all transitions from the ground configuration. The curve with short dashed lines and open squares represents results calculated in the fine structure level mode using all PWB cross sections. The curve with circles was calculated in the fine structure mode using the DW for transitions from the ground configuration. Results from ADPAK are represented by the dashed curve with diamonds. At low temperatures there are large differences among the different approximations. In particular, the two configuration average calculations differ substantially from the two fine structure level calculations. This is because of the metastable states in the Be-like configurations of C III.

Between 10 and 30 eV the calculations are quite close. This is because in this region the radiation is dominated by the Li-like ion C IV. The Li-like ion does not have splitting of singlet and triplet states in the singly excited mode and thus there is essentially no difference in the configuration average and fine structure calculations. At higher temperatures the main differences are due to differences in the PWB and DW cross sections from ground states and the metastable states in He-like C V.

Figures 2-6 show calculated steady-state collisional-radiative power loss rates for Be, B, C, Ne, and Ar versus electron temperature for different densities. For the Be, B, and C calculations the fine structure mode was used with DW cross sections from ground configurations. For Ne and Ar the configuration mode was used with PWB cross sections. For the lighter elements there is a marked decrease in radiated power with increasing density. It should be noted that this is the power per ion per electron. The decrease is due to the increasing collisional deexcitation of excited states before they have time to radiate. The effect diminishes with higher nuclear charge because of faster time scales for radiation. Also included on the figures are results from the ADPAK code[1, 2]. The ADPAK results were calculated at an electron density of $10^{12}$ cm$^{-3}$, but the ADPAK results are essentially independent of density. It can be seen from the figures that the ADPAK results generally follow the detailed calculations for low density. Differences of factors of two are not uncommon, with larger differences at the lower temperatures.

The radiation efficiencies for Be, B, C, Ne, and Ar are presented in figures 7-11. While the results are generally consistent with the results reported by Lengyel[3] and Lackner et al[4], there are quantitative differences. In addition to changes in the radiation rates, the strong density dependence for Be affects the radiation efficiency for temperatures below 30 eV. Boron is affected below ~ 50 eV. C is only affected for very high electron densities (~ $10^{16}$ m$^{-3}$).

IV Conclusions

The calculations of radiated power loss for Be, B, and C presented here are the most comprehensive such calculations we have performed to date. The explicit inclusion of metastable states and use of DW cross sections has been shown to make large differences to radiated power loss, especially at low temperature. Electron density causes large decreases in radiated power loss for light elements with the effect diminishing with increasing atomic number.

For the power levels in present and future tokamaks, this analysis shows that, for the assumed conditions of d ~ 1 cm, ln L ≈ 10, etc., only Neon and Argon can radiate the lowest expected heating powers for DIII-D and ITER and that none of them can handle the highest expected powers. The maximum impurity fraction is limited to 1/Z at which point the plasma is composed of 100% impurities and by the requirement that the central impurity radiation losses be less than the alpha particle heating rate. The dependence of the electron thermal conductivity, κ, on $Z_{eff}$ ($\kappa \sim 1/Z_{eff}$) also decreases the allowed impurity fraction by another factor of $1/Z$, i.e. $F_{Z-max} \sim 1/Z^2$. Using the calculated radiation efficiencies (Figures 7-11), the values of $F/F_{max}$ (Table 2) for $n_s \sim 10^{20}$ m$^{-3}$ indicate that only Ne and Ar (with $f_Z \sim 0.28\%$, $Z_{eff} \sim 1.28$, $f_Z \sim 0.29\%$ with $Z_{eff} \sim 1.94$) for the lowest power DIII-D case and Ar (with $f_Z \sim 0.47\%$, $Z_{eff} \sim 2.52$) for 1.5 GW ITER operation are able to radiate all of the heating power from the edge plasma. This concentration of Argon would radiate almost all of the energy from the plasma core so that ignition almost impossible. Other impurities and higher powers require that $F/F_{max}>1$. Since $F_Z$ scales as $Q_{\parallel}{}^2$, increasing the impurity fraction to increase the radiation losses scales unfavorably. It is also interesting to note that Argon is less effective than Ne for the low temperature DIII-D cases because $1/Z^2$ decreases faster than the radiation efficiency increases with Z. Argon becomes more effective than Ne at higher temperatures where it radiates much more strongly than Ne.

$$F_z(f_Z) \equiv \frac{f_Z(\%)}{1+.01 f_Z(\%) Z(Z-1)} = \left( \frac{\Delta Q_{\parallel}}{n_{es}\sqrt{2\kappa_o T_{es}^2 \int_0^{T_{es}} L_Z(T_e) T_e^{0.5} dT_e}} \right)^2$$

$$f_Z(\%) = \frac{1}{\left( \dfrac{n_{es}\sqrt{2\kappa_o T_{es}^2 \int_0^{T_{es}} L_Z(T_e) T_e^{0.5} dT_e}}{\Delta Q_{\parallel}} \right)^2 - .01\, Z\,(Z-1)} \tag{3}$$

This assessment also indicates that if impurity radiation is to be strong enough to radiate the highest powers incident on a divertor in DIII-D and ITER, it will be necessary to enhance the radiation efficiency by increasing the impurity emissivity by such effects as charge exchange recombination [11, 12] and rapid transport and recycling of impurities as discussed by S. Allen, et al[13] and many others. Such recycling may be possible with gaseous impurities. This enhancement is, however, only effective for $T_e > 10\text{-}20$ eV [11, 13]. For lower temperatures, the present model should be reasonably accurate. The accurate calculation of these effects will require cross sections and rates for charge exchange recombination valid for low energies as well as the calculation of the transport of individual impurity charge states. This, in turn, requires tables of the ionization, recombination, and excitation loss rates for each charge state as a function of temperature and density. Except for charge exchange recombination, these are available, in principle, from our calculations as well. The density dependent effects are important for low power levels.

Table 2  Normalized impurity function $\dfrac{F_Z(f_Z(\%))}{F_Z(f_{Z-max}(\%))} = F_Z(f_Z)\left(\dfrac{100}{Z^2}\right)^{-1}$, required to radiate the

heating power for DIII-D and ITER for $\delta \sim 1$ cm and $n_s. = 10^{14}$ cm$^{-3}$

| | Power MW | $Q_{\|}$ GW/m$^2$ | $T_{sep}$ (eV) | Be Z=4 | B Z=5 | C Z=6 | Ne Z=10 | Ar Z=18 |
|---|---|---|---|---|---|---|---|---|
| fatal fraction (%)[10] | | | | 14 | 9.5 | 6.6 | 2.35 | 0.54 |
| $f_Z(\%)=1/Z$ | | | | 25 | 20 | 16.7 | 10 | 5.56 |
| $F_{Z\,max}$ | | | | 6.25 | 4.0 | 2.8 | 1.0 | 0.31 |
| | | | | $\dfrac{F_Z(f_Z(\%))}{F_Z(f_{Z-max}(\%))}$ | | | | |
| DIII-D 5 MW | 5 | 0.12 | 80 | 4.38 | 2.11 | 1.31 | 0.22 | 0.50 |
| DIII-D 20 MW | 20 | 0.47 | 120 | 22.48 | 10.70 | 6.52 | 1.39 | 1.93 |
| ITER 1.5 GW | 240 | 1.5 | 260 | 26.78 | 11.71 | 6.88 | 1.86 | 0.62 |
| ITER 3 GW | 480 | 3 | 315 | 63.82 | 38.00 | 22.38 | 6.00 | 1.95 |

ACKNOWLEDGMENTS:

   The authors are grateful for encouragement and discussions with N. Putvinski, F. Perkins, R. Hulse, S. Allen, S. Cohen, M. Rosenbluth, G. Janeschitz and K. Lackner.

FIGURES:

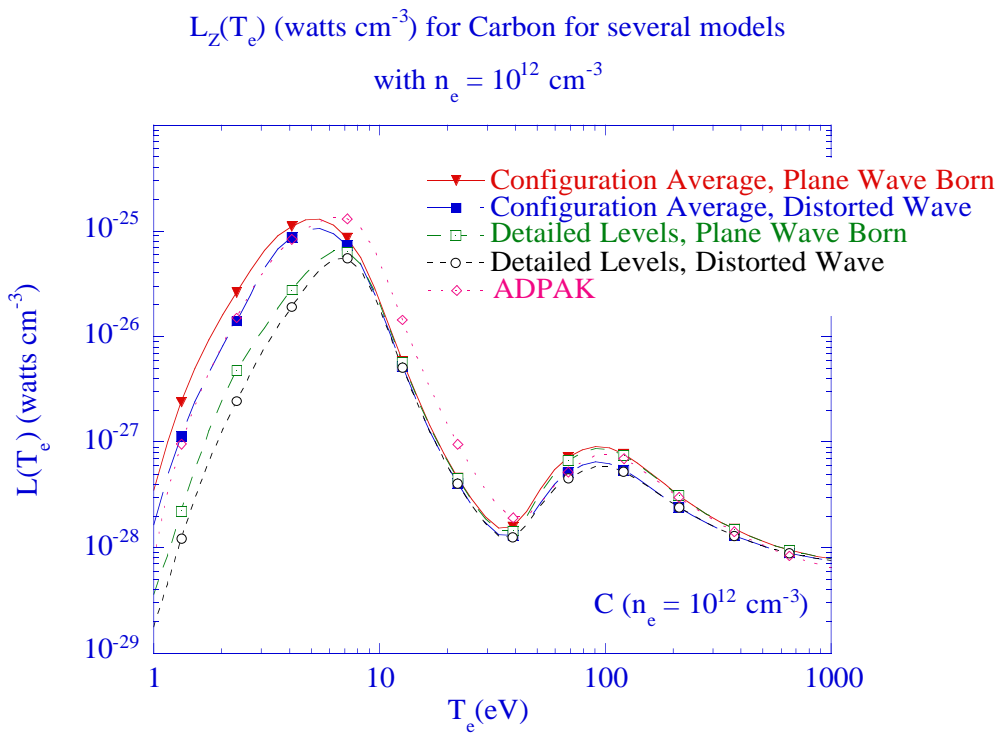

Figure 1. Comparison of Carbon Emission Rates for four models with $n_e = 10^{12}$ cm$^{-3}$.

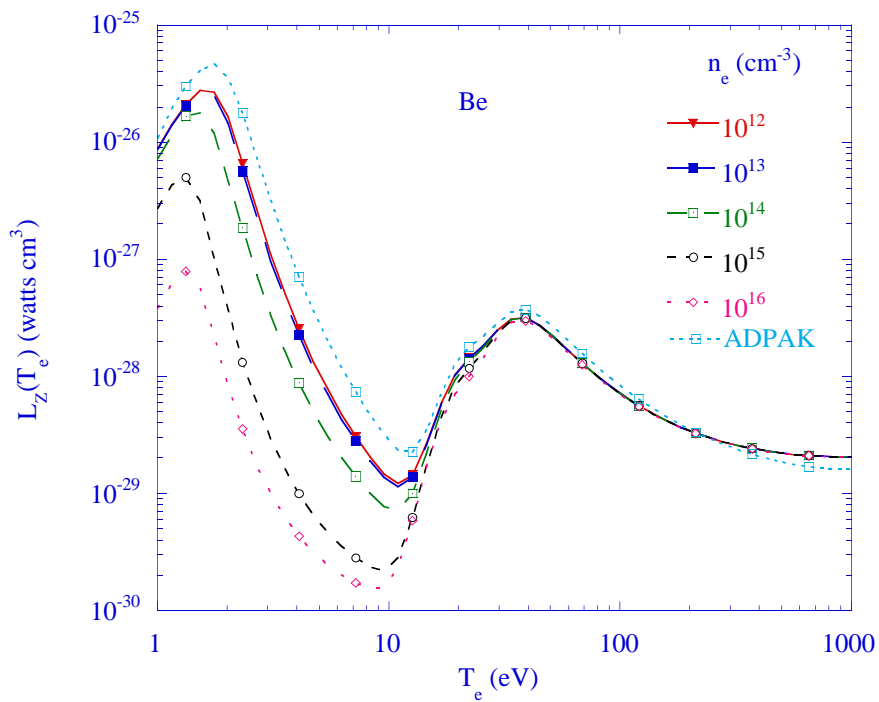

Figure 2. Collisional-Radiative model emission rates for Be for $n_e = 10^{12}$—$10^{16}$ cm$^{-3}$

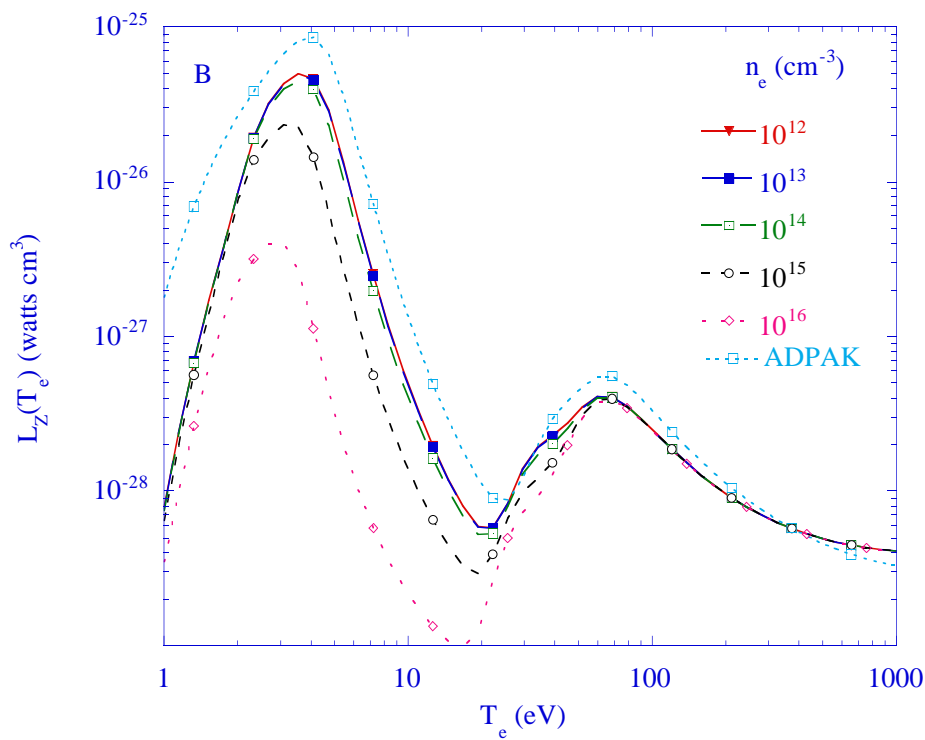

Figure 3. Collisional-Radiative model emission rates for B for $n_e = 10^{12} - 10^{16}$ cm$^{-3}$

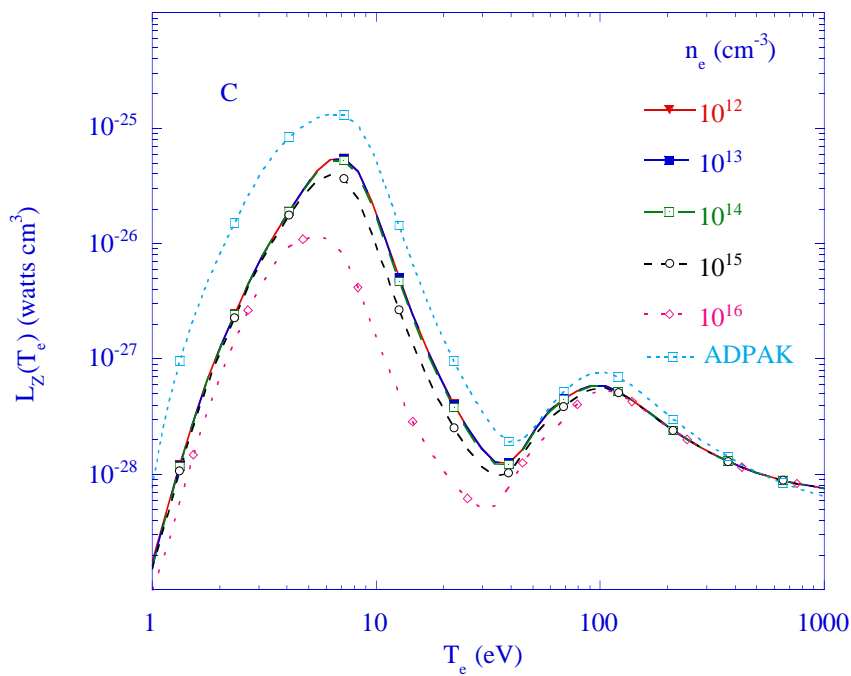

Figure 4. Collisional-Radiative model emission rates for C for $n_e = 10^{12} - 10^{16}$ cm$^{-3}$

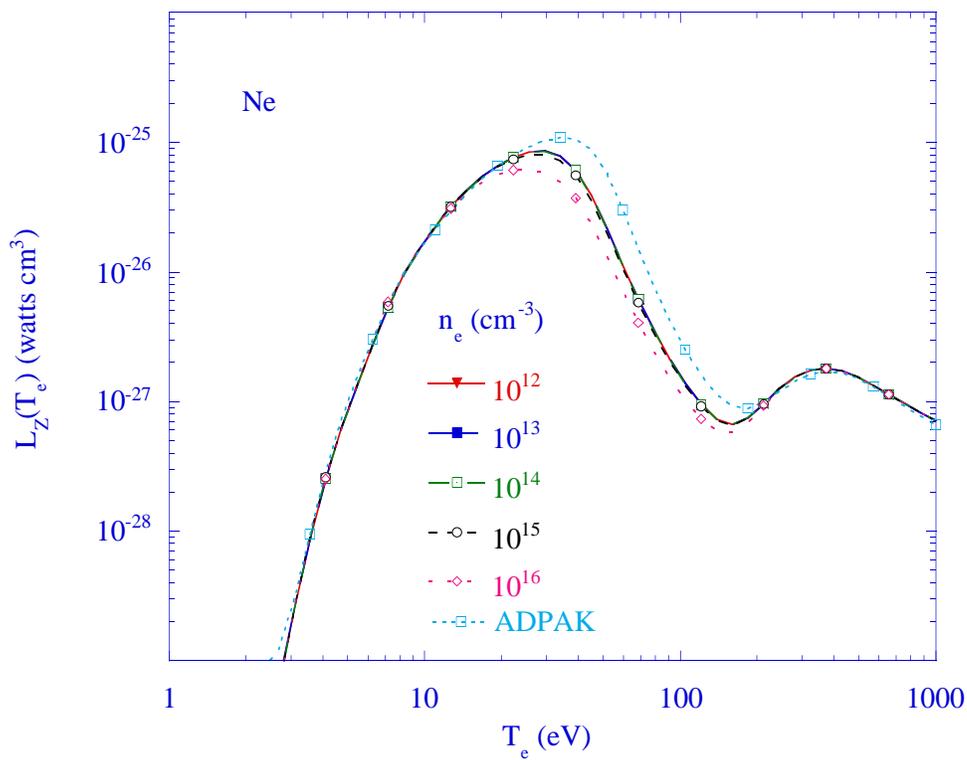

Figure 5. Collisional-Radiative model emission rates for Ne for $n_e = 10^{12} - 10^{16}$ cm$^{-3}$

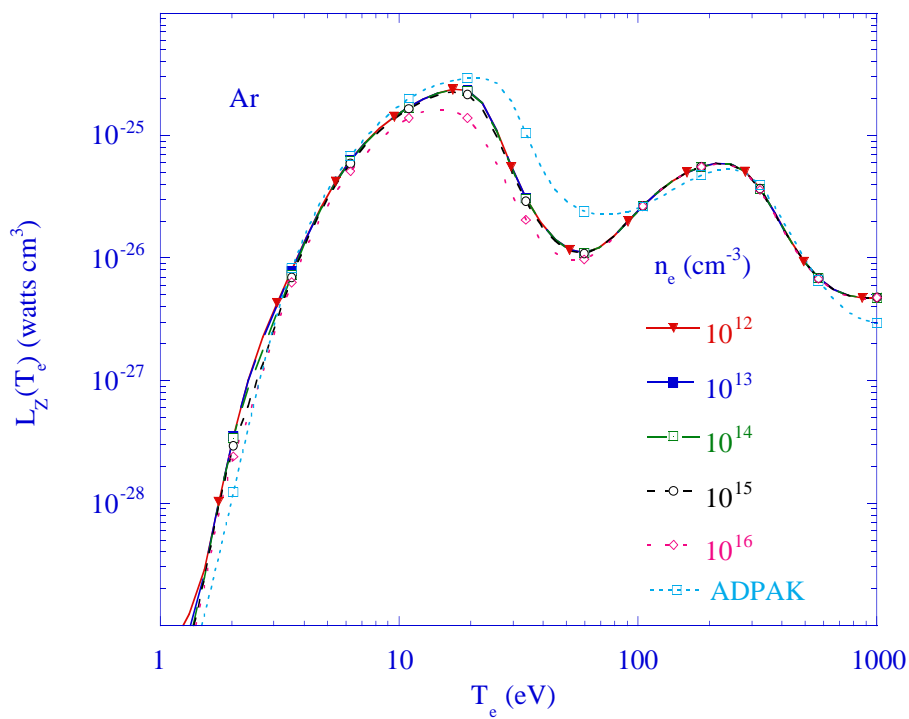

Figure 6. Collisional-Radiative model emission rates for Ar for $n_e = 10^{12} - 10^{16}$ cm$^{-3}$

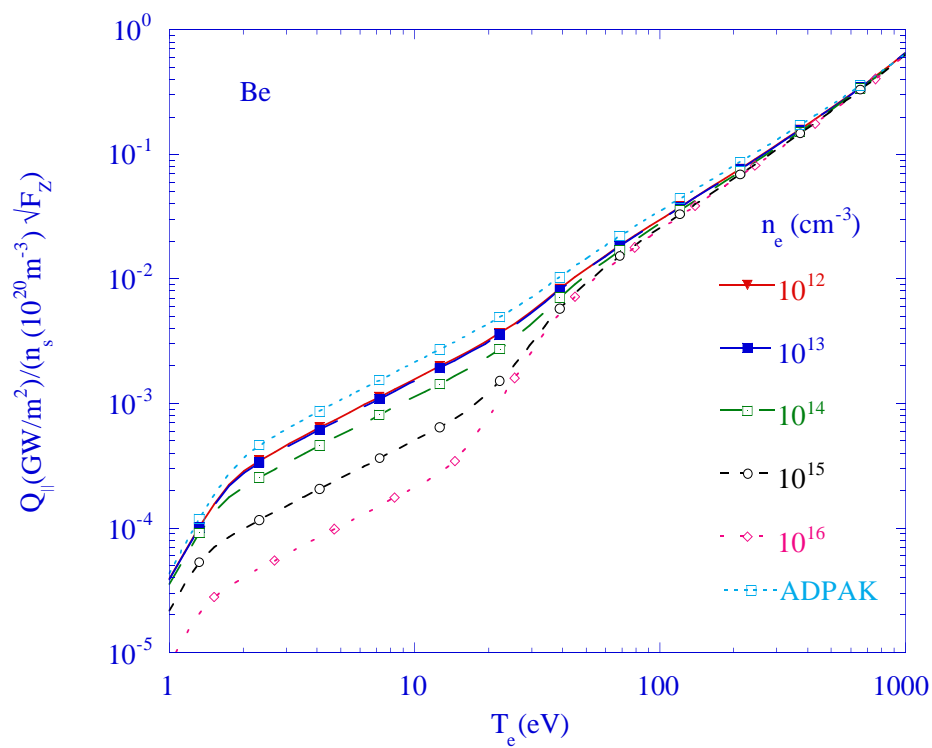

Figure 7. Radiation efficiencies for Be for $n_e = 10^{12}$—$10^{16}$ cm$^{-3}$

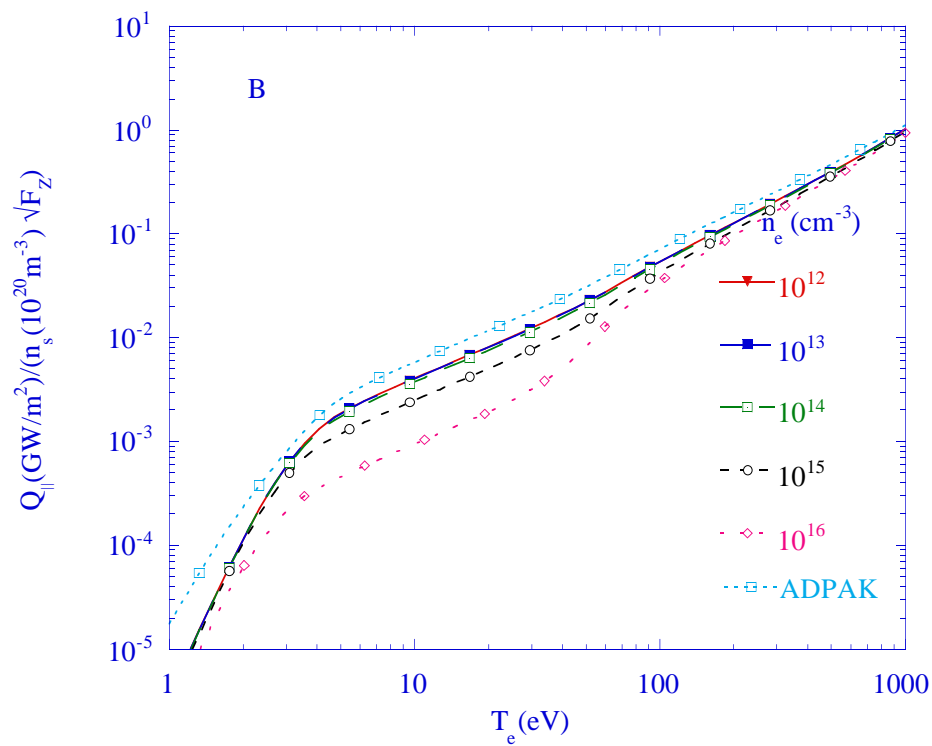

Figure 8. Radiation efficiencies for B for $n_e = 10^{12}$—$10^{16}$ cm$^{-3}$

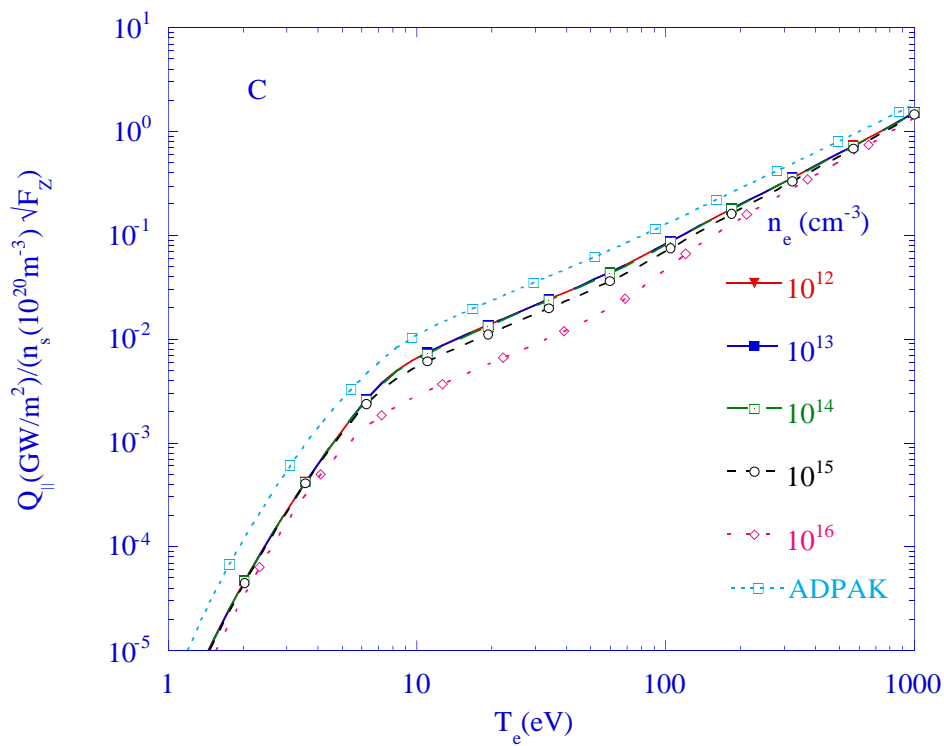

Figure 9. Radiation efficiencies for C for $n_e = 10^{12}$—$10^{16}$ cm$^{-3}$

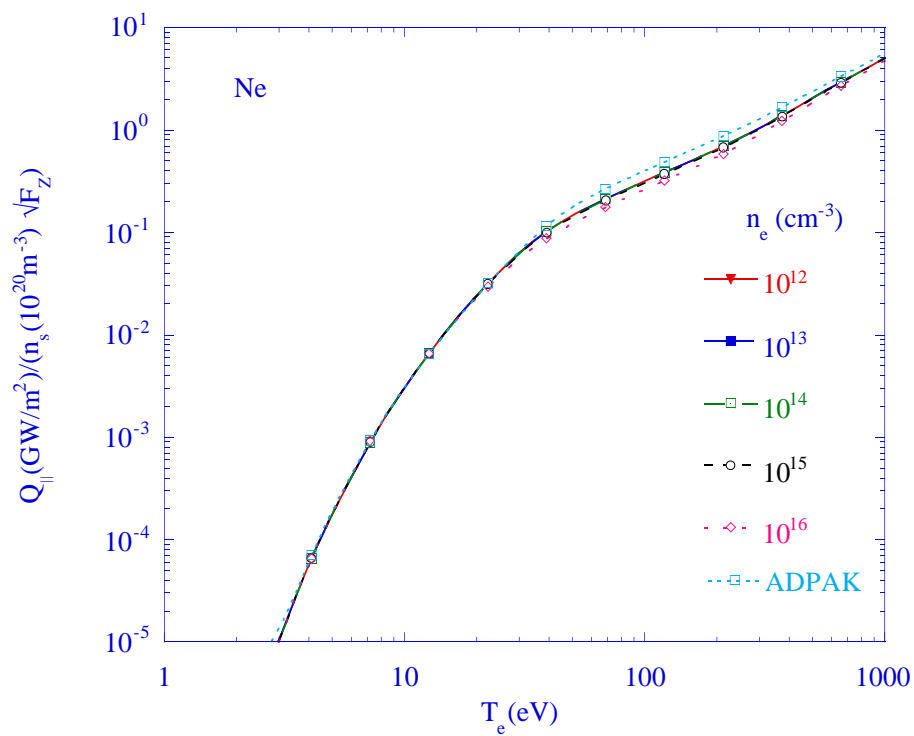

Figure 10. Radiation efficiencies for Ne for $n_e = 10^{12}$—$10^{16}$ cm$^{-3}$

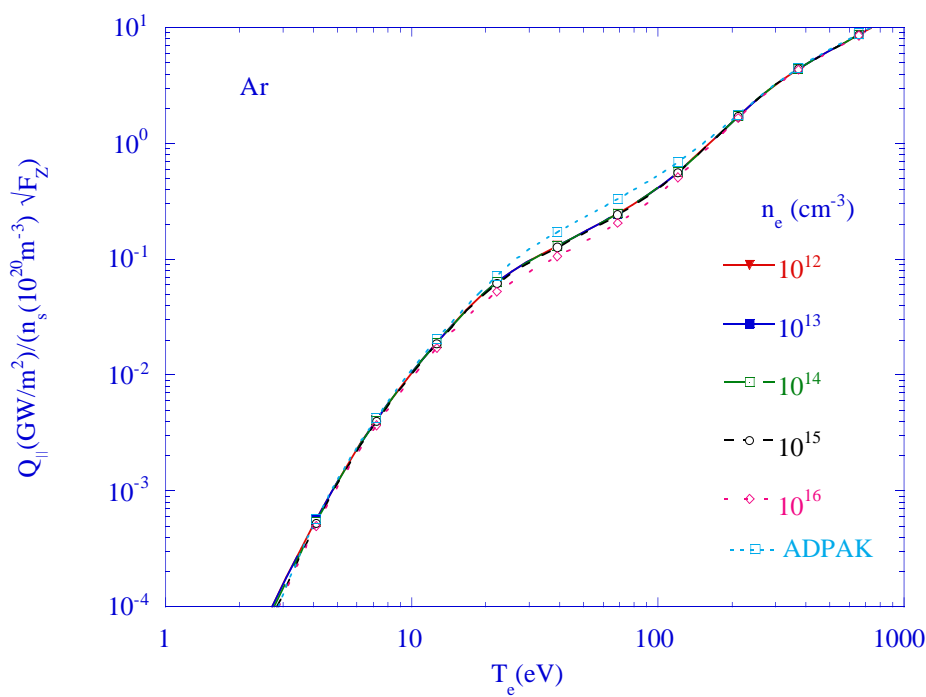

Figure 11. Radiation efficiencies for Ar for $n_e = 10^{12}$—$10^{16}$ cm$^{-3}$